# Theory and computation of electromagnetic transition matrix elements in the continuous spectrum of atoms

### Yannis Komninos[1], Theodoros Mercouris[2] and Cleanthes A. Nicolaides[3]

*Theoretical and Physical Chemistry Institute, National Hellenic Research Foundation,*

*48 Vasileos Constantinou Avenue, Athens 11635, Greece*

e-mail: (1) ykomn@eie.gr, (2) thmerc@eie.gr, (3) caan@eie.gr

## Abstract

The present study examines the mathematical properties of the *free-free* ( *f-f*) matrix elements of the *full* electric field operator, $\boldsymbol{O}_E(\kappa, \vec{r})$, of the *multipolar Hamiltonian*. $\kappa$ is the photon wavenumber. Special methods are developed and applied for their computation, for the general case where the scattering wavefunctions are calculated numerically in the potential of the term-dependent (N-1) electron core, and are energy-normalized. It is found that, on the energy axis, the *f-f* matrix elements of $\boldsymbol{O}_E(\kappa, \vec{r})$ have singularities of first order, i.e., as $\varepsilon' \to \varepsilon$, they behave as $(\varepsilon - \varepsilon')^{-1}$. The numerical applications are for *f-f* transitions in Hydrogen and Neon, obeying electric *dipole* and *quadrupole* selection rules. In the limit $\kappa = 0$, $\boldsymbol{O}_E(\kappa, \vec{r})$ reduces to the *length* form of the electric dipole approximation (EDA). It is found that the results for the EDA agree with those of $\boldsymbol{O}_E(\kappa, \vec{r})$, with the exception of a wave-number region $k' = k \pm \kappa$ about the point $k' = k$.

## I. INTRODUCTION

The present work has been carried out in the context of our studies of the nonperturbative solution, $\Psi(t)$, of the *many-electron time-dependent Schrödinger equation* (METDSE), $\boldsymbol{H}(t)\Psi(t) = i\hbar \dfrac{\partial \Psi(t)}{\partial t}$, describing the interaction of atoms and molecules with strong and/or ultrashort electromagnetic pulses, which is achieved via the *state-specific expansion approach* (SSEA), whose features, methodology and characteristic applications were reviewed recently [1,2].

The physical situations that the SSEA covers, (which are numerous), are assumed to be described by a total Hamiltonian (for an atom or a molecule) of the form $H(t) = \mathbf{H}_{atom(molecule)} + V(t)$. Once the interaction $V(t)$ has been chosen, the fundamental requirement is the possibility of computing reliably those *state-specific* stationary wavefunctions for the discrete and the continuous spectrum of field-free Hamiltonian $\mathbf{H}_{atom}$ (we drop the suffix for molecules) which are the most relevant to the quantitative solution of the problem under consideration.

The SSEA follows from the fundamental quantum mechanical principle of the expansion of a wavepacket in terms of the complete set of N-electron stationary states. After an analysis of the apparent requirements in terms of the pulse characteristics and the spectrum of the stationary states, the SSEA solution, $|\Psi(t)>_{SSEA}$, is constructed and computed in the form, (we omit the index for each possible channel),

$$|\Psi(t)>_{SSEA} = \sum_m a_m(t)|m> + \int_0^\infty b_\varepsilon(t)|\varepsilon> d\varepsilon \qquad (1)$$

In the *state-specific* expansion (1), $|m>$ represents the state-specific bound wavefunctions that are relevant to the situation, and $|\varepsilon>$ are the energy- normalized (Dirac normalization) scattering states. Substitution of eq. (1) into the METDSE leads to integro-differential coupled equations, whose solution yields the complex, time-dependent mixing coefficients. Once the bound wavefunctions, (for the discrete states and for the localized component of resonance states), and the scattering wavefunctions have been obtained, one needs to calculate 'bound-bound', (b-b), 'bound-free', (b-f), and 'free-free' (f-f) matrix elements of $V(t)$.

In the present work, $V(t)$ is chosen to be the interaction from the *multipolar* Hamiltonian, $H_{MP}(t)$, rather than from the *minimal- coupling* Hamiltonian, $H_{MC}(t)$ [3, 4]. The reasons and the results are explained and presented in the following paragraphs and sections.

As the title indicates, the analysis and methods of computation deal with the calculation of matrix elements between energy-normalized scattering wavefunctions, $<\varepsilon\ell|H_{MP}(t)|\varepsilon'\ell'>$, *on- and, especially, off-resonance,* with respect to the photon energy $\hbar\omega$. The scattering orbitals, $\varepsilon\ell$, are calculated *numerically* in the term-dependent multi-electron core of each channel, and have the correct asymptotic boundary conditions.

We stress that the theory is formulated around the concepts of parity and of angular and spin selection rules, in spherical symmetry, and not around the standard multipole expansion which results from the Taylor-series of $e^{i\vec{K}\cdot\vec{r}}$ in the matter-radiation interaction formulas. It is clear that, in contradistinction to the textbook argument (going back to Dirac) of the validity of the lowest-order term based on 'atomic dimensions' and the *long-wavelength approximation* (LWA), when it comes to matrix elements between the unbound scattering wavefunctions this argument cannot be justified a priori.

Specifically, first a particular combination is chosen for initial and final scattering atomic states, $| E\ell >$ and $| E'\ell' >$, ( $E\ell = E_{core}(^{2S+1}L) + \varepsilon\ell$ ), and then the result for the corresponding matrix element using $\boldsymbol{H}_{MP}(t)$ is obtained. The main mathematical and computational difficulties, having to do with the oscillatory behavior of the scattering functions and with the presence of singularities in the integrals, come from the fact that the state-specific, channel-dependent, scattering orbitals are energy-normalized and are obtained *numerically*. The latter fact is necessary for the treatment of many-electron atoms of arbitrary symmetries and electronic structures, with one or more open channels, as it has been proposed and implemented in the SSEA [1, 2].

We restrict the demonstration to two cases:

1) Parity changes and $\Delta j = \pm 1$, corresponding to *electric dipole plus higher order contributions*. By 'higher order' we mean contributions which, in the standard multipole expansion using $\boldsymbol{H}_{MC}(t)$, correspond to higher multipoles satisfying the same selection rules, e.g., electric octopole with $\Delta j = \pm 1$.

2) Parity does not change and $\Delta j = \pm 2$, corresponding to *electric quadrupole plus higher order contributions*.

Atomic units are used throughout the paper $(\hbar = e = m_e = 1)$. The speed of light in these units is $c = 137.037$ a.u.

## II. SINGULARITIES IN THE *FREE-FREE* MATRIX ELEMENTS

Once the coupled equations resulting from the expansion (1) have been constructed in terms of matrix elements containing *numerical*, energy-normalized scattering wavefunctions, the number of *f-f* matrix elements that must be calculated ranges from large (in the tens of thousands) to huge (in the hundreds of thousands and beyond), depending on the problem. For a calculation to be reliable, this fact implies that any possible systematic inaccuracies in the numerical calculation of the energy integrals containing *f-f* matrix elements must be eliminated rigorously. In principle, a source of such inaccuracies is the presence of *singularities* for energy differences *off-resonance* with the photon energy. The discovery, the mathematical understanding and the proper numerical handling of these singularities when the wavefunctions are *numerical*, have been one of the objects of our research program on the non-perturbative solution of the METDSE.

In the initial formulation and implementation of the SSEA [5], the coupling operator $V(t)$ in the matrix elements had the '*velocity*' form, $\vec{\nabla}$. This is the form to which the matrix elements of $H_{MC}(t)$ are reduced in the EDA. The method for calculating numerically the *f-f* matrix elements for energy-normalized scattering wavefunctions was described in the text and in the Appendix A of [5]. The main concern was the proper handling of the singularity which exists for $\varepsilon = \varepsilon'$, where $\varepsilon$ is the energy of the free electron. The *f-f* matrix element of the *velocity* operator has a pole of first order at $\varepsilon = \varepsilon'$, whereas that of the *length* form has a pole of second order, as is evident from the commutator relating the two operators. The problem of the '*diagonal*', ($\varepsilon = \varepsilon'$), (or, '*on-shell*') singularity in *f-f* matrix elements within the *length* EDA had been discussed and treated carefully in terms of the *analytic* hydrogen functions in the late 1980s, by Madajczyk and Trippenbach [6] and by Veniard and Piraux [7].

In the mid-1990s, our exploration of the nature of the diagonal singularities in *f-f* matrix elements and of their significance in the problem of solving the TDSE non-perturbatively, continued in [8]. We focused on two issues: The first had to do with the degree of rigor, reliability and efficiency of box-normalized basis sets in terms of which $H_{atom}$ can be diagonalized, (this is practical and reliable only for special two-electron systems), to produce wavefunctions that are expected to represent approximately the states of both the discrete and the continuous spectrum and used in the expansion of $\Psi(t)$. The second had to do with the mathematical properties of the two interaction operators comprising $\boldsymbol{H}_{MC}(t)$, (without the EDA), namely, $\vec{A}(\vec{r},t)\bullet\vec{p}$ and $\vec{A}^2(\vec{r},t)/2$ [8].

As regards the second issue, which is linked to the topic of interest here, we showed that, whereas the *f-f* matrix element of $\vec{A}(\vec{r},t)\bullet\vec{p}$ is free of singularities, the second part, $\vec{A}^2(\vec{r},t)/2$, and therefore the full $\boldsymbol{H}_{MC}(t)$, gives rise to a logarithmic singularity for selection rules of the electric dipole as well as of the higher electric and magnetic terms [8]. On the other hand, the singularity exhibited by the *f-f* matrix elements of the $\boldsymbol{H}_{MP}(t)$ Hamiltonian is of the first order.

The above results might lead one to expect that, if the great leap to a future attempt of developing and applying theory and methodology for solving the METDSE by an expansion method with real scattering functions using the full Hamiltonian were to be taken, $\boldsymbol{H}_{MC}(t)$ ought to be the appropriate choice. Indeed, the familiarity with the vector potential and with $\boldsymbol{H}_{MC}(t)$ in the theory of atomic spectroscopy and of quantum optics, goes back to the first decade of quantum mechanics, e.g., [9].

However, apart from the awesome computational requirements, there is also the issue of the interpretation of the expansion coefficients as probability amplitudes. This problem can be solved formally at the level of the EDA by applying the Pauli gauge transformation on the solution, which holds for all *t* [10-12]. In computational practice, one may choose the points in time, $t'$, when $\vec{A}(\vec{r},t') = 0$, such as when the duration of the pulse is over. When it comes to the complete $\boldsymbol{H}_{MC}(t)$, it is not clear how the formal gauge transformation can be applied in a practical way. We surmise that the above argument about the special points in time can still be useful.

The notion of using the full atom-radiation interaction confronted us when we tackled the problem of obtaining the time-dependent probabilities for excitation of Rydberg states (wavepackets) [13,14]. For reasons explained again in the review [1], rather than dealing with $\boldsymbol{H}_{MC}(t)$ we turned to the theory of $\boldsymbol{H}_{MP}(t)$ and developed a formalism that is adapted to the calculation of matrix elements with atomic wavefunctions.

### III.  OUTLINE OF THE THEORY AND RESULTS OF THE FOLLOWING SECTIONS

We focus on the structure and computation of the *f-f* matrix elements of the time-independent full electric operator, $\boldsymbol{O}_E(\kappa, \vec{r})$ [3, 4, 13], defined in eq. (4) of the following section, when the free electron energies are *on-* and, especially, *off-resonance*, $(\varepsilon - \varepsilon' \neq \hbar\omega)$, with the photon energy, $\hbar\omega$.  In particular, we present results for $< \varepsilon\ell \,|\, \boldsymbol{O} \,|\, \varepsilon'\ell' >$, for the following cases:

1) $\boldsymbol{O}$ is the *length* EDA operator, $\vec{r}$.  The electric dipole selection rule applies.

2) $\boldsymbol{O}$ is $\boldsymbol{O}_E(\kappa, \vec{r})$.  By applying the dipole (E1) and the quadrupole (E2) selection rules, we explore the mathematical properties of the matrix element, identify the nature of the singularities, and compare the results to those obtained from the *length* EDA. (We recall that $\boldsymbol{O}_E(\kappa, \vec{r})$ reduces to the *length* form in the EDA).

In the numerical application, we use a particular example from a generic pump-probe arrangement: The initial $| \varepsilon\ell >$ with which the probe pulse of photon energy $\hbar\omega$ interacts, is assumed known from, say, a photoionization process (pump step). Then, this is allowed to couple with a probe pulse at various values of frequency. The matrix elements of $\boldsymbol{O}_E(\kappa, \vec{r})$ between $| \varepsilon\ell >$ and other scattering orbitals, $| \varepsilon'\ell' >$, are evaluated by first choosing the E1 and E2 selection rules mentioned in the Introduction. The formalism is general and allows the consideration of higher electric selection rules.

It also allows the treatment of the case of the full magnetic operator, for which the fundamentals have already been published, e.g., [1] and references therein. However, given the requirement of economy, this case is not examined here.

The example that we chose for numerical study assumes the photoionization of the *2s* electron of Neon, as in [15]. For our present computations of matrix elements, $| \varepsilon\ell >$ is taken to be the scattering orbital $| \varepsilon p >$, $\varepsilon = 13.6$ eV, corresponding  to the photoionization of the *2s* orbital of the Neon ground state, *Ne* $1s^2 2s^2 2p^6 \ ^1S$, by a pump photon of energy 62 eV.  According to the above, for a probe pulse with linear polarization, the full operator matrix elements of interest are $< \varepsilon p \,|\, \boldsymbol{O}_E(\kappa, \vec{r}) \,|\, \varepsilon's >$ or $\varepsilon'd >$ for E1, $\Delta j = \pm 1$, selection rules, and $< \varepsilon p \,|\, \boldsymbol{O}_E(\kappa, \vec{r}) \,|\, \varepsilon'f >$ for E2, $\Delta j = \pm 2$, selection rules. All scattering orbitals are

computed numerically in the potential of the $Ne^+$ $1s^2 2s 2p^6$ $^2S$ core. The energy range for the electron in the final state is taken to be, $\Delta\varepsilon' = [8.7 - 19.6]$ eV. The frequency of the probe photon is 74.6 eV (XUV region). By choosing the XUV rather than the IR region, the range of wavenumbers where the discrepancy between the results from the EDA and from $\boldsymbol{O}_E(\kappa, \vec{r})$ is significant, is larger. This statement follows from the results and analysis of this work.

For the purpose of the present formalism, it is more convenient to use wavenumbers rather than energies. The standard dispersion relations are, $\varepsilon = \frac{1}{2}k^2$ for electrons and $\omega = \kappa c$ for photons.

## IV. ELECTRIC FIELD OPERATOR

According to Loudon [3], equation 5.35, the electric part of the *multipolar* Hamiltonian can be written elegantly as,

$$H_{el} = e \sum_j \int_0^1 \vec{r}_j \, \vec{E}(\lambda \vec{\kappa} \cdot \vec{r}_j) d\lambda \tag{2}$$

where $\vec{\kappa}$ is the wave-vector and $\vec{E}$ is the electric field. Unlike the usual *minimal coupling* Hamiltonian, it contains the fields instead of the potentials. This makes it gauge-independent. Thus, as discussed in the previous section, in a perturbation treatment with (2) as the interaction Hamiltonian, the expansion coefficients are interpreted as probability amplitudes. The two Hamiltonians are connected through a unitary transformation [3, 4].

With $\vec{\varepsilon}$ as the polarization vector, $\vec{E}$ is written as,

$$\vec{E} = \frac{1}{2} E_0(t) \, \vec{\varepsilon} \, e^{i\vec{\kappa}\vec{r} - i\omega t} + c.c. \tag{3}$$

The simplest case arises when the $z$-axis is chosen in the direction of the wave-vector $\vec{\kappa}$. Then, by convention, the polarization $\vec{\varepsilon}$ of the electric field is along the x-axis.

Under this choice, the inner products in the above expressions assume the forms $\vec{\kappa}\vec{r} = \kappa r \cos\theta$ and $\vec{\varepsilon}\vec{r} = \varepsilon r \sin\theta \cos\phi$. Then, the general formula (2) can be reduced, in the case of atoms, to an expression in terms of integrals of spherical

Bessel functions multiplied by an angular part containing spherical harmonics [13] - see the Appendix. The result is that $H_{el}$ can be written in the form,

$$H_{el} = \frac{1}{2} E_0(t) e^{-i\omega t} \sum_{\ell=1}^{\infty} i^{\ell+1} (2\ell+1) F_\ell(\kappa r) \Theta_\ell(\theta, \phi) + c.c.$$

$$\equiv \frac{1}{2} E_0(t) e^{-i\omega t} \boldsymbol{O}_{E(\kappa, \vec{r})} + c.c. \qquad (4)$$

where $E_0(t)$ is the amplitude of the electric field with wave-vector $\kappa$,

$$F_\ell(\kappa r) = \frac{1}{\kappa} \int_0^r \frac{1}{r'} j_\ell(\kappa r') dr', \quad j_\ell(\kappa r') \text{ is the spherical Bessel function.} \qquad (5)$$

and

$$\Theta_\ell(\theta, \phi) = \sqrt{\frac{\pi \ell(\ell+1)}{2\ell+1}} \left( Y_\ell^{-1} - Y_\ell^1 \right) \qquad (6)$$

$Y_\ell^1$ is the spherical harmonic. The result of the angular integration over the spherical harmonics of the initial and the final state $< Y_{\ell_1 m_1} | \Theta_\ell(\theta, \phi) | Y_{\ell_2 m_2} >$ can be expressed in terms of $3j$ symbols [16,17]. According to the rules of the 3-j symbols, the angular part causes transitions with $m_2 = m_1 \pm 1$ and $|\ell_1 - \ell_2| \leq \ell \leq \ell_1 + \ell_2$, whilst $\ell_1 + \ell_2 + \ell = even$. Consequently, the infinite summation over $\ell$ is reduced to a finite one. For transitions with $|\ell_1 - \ell_2| = 1$, (EDA selection rule), $\ell$ acquires odd values. For transitions with $|\ell_1 - \ell_2| = 2$, (electric quadrupole approximation (EQA) selection rule), $\ell$ acquires even values.

The spherical Bessel functions of the integrand can be expressed as infinite polynomials. For small values of the argument $\kappa r$, we may keep, to a good approximation, the first few powers only. This is the LWA, whose use in quantum mechanics is extensive and pervades many fields, especially in the form of the EDA and of the EQA, where only the smallest power of $\kappa r$ in the expansion is kept. This is best seen as the result of taking the limit $\kappa \to 0$.

Since

$$\lim_{\kappa \to 0} j_\ell(\kappa r) = \frac{(\kappa r)^\ell}{(2\ell+1)!!},$$

one obtains

$$(2\ell+1)\lim_{\kappa\to 0}F_\ell(\kappa r) = \kappa^{\ell-1}\frac{r^\ell}{\ell(2\ell-1)!!}.$$

For EDA, the first term, $\ell = 1$, of the expansion with odd values of $\ell$, gives

$$3F_1 \approx r.$$

For EQA, the first term, $\ell = 2$, of the expansion with even values of $\ell$, gives

$$5F_2 \approx \frac{1}{6}\kappa r^2.$$

The appearance of $\kappa$ makes the matrix element of the quadrupole term much smaller than the dipole one. This is due to the fact that $\kappa = \omega/c$, where $c = 137.037$ while $\omega$ is of the order of unity for XUV energies. The same is true for the magnetic dipole term. Therefore, unless hard X-rays are involved, the dipole is the dominant term.

Higher-order multipoles get contributions from more than one terms. For example, the octopole results from the second term in the expansion of $j_1(\kappa r)$ and the first term of $j_3(\kappa r)$.

At this point, an important distinction must be made between applying the EDA and applying electric dipole selection rules. In the latter case, the inclusion of the higher spherical Bessel functions in the calculation depends on the angular momenta of the initial and final states. For example, $s \to p$ transitions involve only $j_1(\kappa r)$, $p \to d$ transitions involve two terms containing $j_1(\kappa r)$ and $j_3(\kappa r)$ etc. The same distinction must be made between applying the EQA and applying quadrupole selection rules.

We close by emphasizing that for free-free matrix elements, the Taylor expansion of $e^{i\vec{k}\vec{r}}$, is totally inconvenient beyond the quadrupole term, $\sim r^2$, since the integrals blow up extremely quickly. On the contrary, in our formalism, the spherical Bessel functions already contain all the higher order terms at each stage of the angular momentum expansion.

## V. THE RADIAL INTEGRALS OF THE FULL ELECTRIC FIELD OPERATOR

As we have shown in the previous section, $\boldsymbol{O}_E(\kappa,\vec{r})$ is expressed in terms of integrals of spherical Bessel functions appearing in eq. (5)

$$I_\ell^{(-1)}(z) \equiv \int_0^z \frac{1}{x} j_\ell(x) dx \tag{7}$$

for $\ell > 0$. The index (-1) refers to the power of $x$. In the following, the parenthesis is dropped.

The integrals $I_\ell^{(1)}$ which appear in the paramagnetic operator have been presented in a previous paper [17]. They are amenable to a similar treatment as $I_\ell^{(-1)}$, and lead to matrix elements of the same type. Hence, for reasons of economy, they will be omitted from this work.

The radial functions $I_\ell^{-1}(\kappa r)$, where $\kappa$ is the wavenumber of the photon, appear in the expression for $\boldsymbol{O}_E(\kappa,\vec{r})$, each one multiplied by the appropriate angular factor. In the case where $\kappa = 0$, we have the result

$$\frac{2\ell+1}{\kappa} I_\ell^{-1}(\kappa r) \to \delta_{\ell 1} r \tag{8}$$

which is the *length* EDA. This result is obtained from eq. (4), by substituting the small argument expression of the spherical Bessel functions.

As it was shown in [17], the integrals are computed by the recursive relation

$$I_\ell^{-1}(z) = \frac{\ell-2}{\ell+1} I_{\ell-2}^{-1}(z) - \frac{2\ell-1}{\ell+1} \frac{j_{\ell-1}(z)}{z} + \frac{1}{3}\delta_{\ell 2} \quad , \qquad \ell \geq 2 \tag{9}$$

The odd-$\ell$ integrals start with the function

$$I_1^{-1}(z) = \frac{1}{2}[Si(z) - j_1(z)], \tag{10a}$$

and the even-$\ell$ ones with the function

$$I_2^{-1}(z) = \frac{1}{3} - \frac{1}{z} j_1(z). \tag{10b}$$

Generally, we may write

$$I_\ell^{-1}(z) \approx \frac{(\ell-2)!!}{(\ell+1)!!}(p+1)I_p^{-1}(z) - \frac{2\ell-1}{\ell+1}\frac{j_{\ell-1}(z)}{z} - \frac{(\ell-2)(2\ell-5)}{(\ell+1)(\ell-1)}\frac{j_{\ell-3}(z)}{z}.... \quad (11)$$

where $p=1$ for the odd-$\ell$ values and $p=0$ for the even-$\ell$ values, putting $I_0^{-1}(z)=1$.

For large values of $r$, these integrals reach a constant value. This can be seen by rearranging eq. (7) as,

$$I_\ell^{-1}(\kappa r) \equiv \int_0^\infty \frac{1}{x}j_\ell(x)dx - \int_{\kappa r}^\infty \frac{1}{x}j_\ell(x)dx \quad (12)$$

The first integral has a constant value. For the even-$\ell$ series this constant is obtained from eq. (10b), whilst for the odd-$\ell$ series it results from the properties of the sine integral function in eq. (10a). As it is stated in the mathematics literature [18],

$$Si(z) \equiv \frac{\pi}{2} + si(z) \quad (13)$$

where $si(\kappa r)$ approaches zero for large values of $r$.

The constant term in the full electric operator is important, since it prevails in the asymptotic region. Consequently, in this region the matrix elements of $\boldsymbol{O}_E(\kappa,\vec{r})$ contain an overlap between scattering states. As shown in [17] and in eq. (15) below, this overlap contains the singularity of the matrix element at equal energies.

In order to stress this fact, we introduce the notation,

$$I_\ell^{-1}(\kappa r) = A_\ell + J_\ell^{-1}(\kappa r), \qquad A_\ell = \frac{(\ell-2)!!}{(\ell+1)!!}d_\ell \quad (14)$$

where $J_\ell^{-1}$ is the part of the function of eq. (5) with the constant term removed, so that $J_\ell^{-1}(z) \to 0$ for large values of $z$. In eq. (14), $d_\ell = \frac{\pi}{2}$ for the odd-$\ell$ values, and $d_\ell = 1$ for the even-$\ell$ values. Note that the constant term is independent of $\kappa$.

The $f$-$f$ matrix elements involve two scattering wavefunctions, $y_{kl}$ and $y_{k'l'}$, which extend to infinity as

$$y_{kl}(r) \to \sqrt{\frac{2}{\pi k}}\sin(kr+\frac{1}{k}\log kr+\delta_{kl}) \quad (15)$$

where

$$\delta_{kl} = \sigma_l(k) + l\frac{\pi}{2} + \delta_{kl}^c \qquad (16)$$

$\sigma_l(k)$ is the Coulomb phase shift and $\delta_{kl}^c$ is the additional one due to core electrons. Obviously, for hydrogen $\delta_{kl}^c$ is zero.

The resulting radial integrals are of the type,

$$\mathcal{E}_{klk'l'}^{\kappa\ell} = \frac{2\ell+1}{\kappa}\int_0^\infty I_\ell^{-1}(\kappa r)y_{kl}(r)y_{k'l'}(r)dr \qquad (17)$$

Choosing a radius $r_0$, we divide the space into an inner and an outer region, where different methods are applied in the evaluation of the integrals (17). In the inner region, the N-electron Hamiltonian describes all the many-body interactions. The energy-normalized scattering wavefunctions are obtained as numerical solutions of a model potential, usually the fixed-core HF one. In the outer region, the potential is effectively hydrogenic and the wavefunctions are computed accurately in a semi-analytic way, using the Wentzel-Kramers-Brillouin (WKB) approximation. The effect of the core orbitals is reflected in the presence of phase shifts.

Below we shall discuss the two regions separately.

### A. The inner region, $r < r_o$

In the inner region, the sine-integral and the spherical Bessel functions in the expression of $I_\ell^{-1}(\kappa r)$ in eq. (11), are calculated through their power series. The scattering functions are computed numerically. Hence, the calculation of the inner part of the integrals (17) is done numerically. In practice, the choice of the radius $r_0$ is guided mainly by two factors: The scattering wavefunctions for $r \geq r_0$ should be described accurately by the WKB form and the continued fraction which is employed for the computation of the function $si(\kappa r)$ in eq. (13), (see, also, Appendices A and B), should converge rapidly in the complex plane.

**B. The outer region, $r \geq r_o$**

The operator of eq. (4) extends to distances $r \geq r_o$, a region where the numerical calculation of integrals containing rapidly oscillating functions becomes highly inaccurate. The way out of this problem is to use the fact that, in the outer region, the scattering functions assume the WKB form.

Since the functions entering the integral (17) extend to infinity, the method of calculation of its outer part has to be more sophisticated than that employed for the inner one. Fortunately, the WKB form of the scattering functions (see Appendix D of [19])

$$y_{kl}(r) = \sqrt{\frac{2}{\pi}} \frac{1}{\sqrt{\zeta_{kl}(r)}} \sin \phi_{kl}(r) \tag{18}$$

gives highly accurate results in this region.

To start with, we examine the evaluation of the outer part of the integral (17) in the extreme case $\kappa = 0$, when the operator assumes the *length* form. In this case, the practical approach is to revert to the calculation of the "*acceleration*" integral,

$$S^{(-2)}_{kl;k'l'}(r_0) \equiv \int_{r_0}^{\infty} \frac{1}{r^2} y_{kl}(r) y_{k'l'}(r) dr \tag{19}$$

which is linearly related to the *length* one (see [20] and [19] and discussion below).

Sil, Crees and Seaton [21] have developed a fast and accurate method to calculate integrals of this type. Specifically, the sine functions in the WKB expression of the orbitals, eq. (18), enter the integral (19) in the complex exponential form, while the resulting integrals are subsequently transformed by deforming the integration contour into the complex plane. As a consequence, the oscillating complex exponential functions become exponentially decreasing ones, and the integrals are calculated efficiently by Gaussian quadrature. Unless very close to threshold, the main contribution results from integrands containing $e^{\pm i(\phi_{kl}(r) - \phi_{k'l'}(r))}$ rather than $e^{\pm i(\phi_{kl}(r) + \phi_{k'l'}(r))}$.

The regularization in this way of unbound integrals in atomic and molecular physics containing oscillatory complex functions is in harmony with the generic procedure that was introduced for the regularization of resonance wavefunctions with complex energies [22,23]. The physical basis of this procedure is the standard notion of separation of the *asymptotic* (*outer*) region that refers to the wavefunction of the free particle outside the effective range of interactions, from the *inner region* where the interactions of all particles must be accounted for correctly. Integration in the inner region is done on the real axis, and integration in the outer region, $r \geq r_0$, is regularized by completing integration in the complex plane. Since 1978 [22], this technique has found practical use in many problems of atomic and molecular physics under the name *exterior complex scaling* (ECS), a term coined by Simon [24], when he applied it to a formal analysis of molecular resonances (without calculations) in the Born-Oppenheimer approximation.

We now proceed to examine the mathematically rather subtle case arising from the outer- region part of the integral (17).

The key observation is that the functions $J_\ell^{-1}(z)$ can be written in the general form $\dfrac{Q_\ell(\kappa r)}{r^2} e^{\pm i \kappa r}$ where $Q_\ell(x)$ is a monotonic analytic function, such that $\lim_{x \to \infty} Q_\ell(x) \to 1$. This is shown in Appendix A. Specifically, the spherical Bessel functions are of this form when written in terms of the corresponding Hankel functions, whilst the function $si(z)$ can be expressed in terms of a continued fraction times an exponential (Appendix B).

Thus, we are led to integrals of the general type,

$$G_{\kappa \ell; kl; k'l'}^{(-2)}(r_0) \;=\; \frac{1}{\kappa^2} \int_{r_0}^{\infty} \frac{Q_\ell(\kappa r)}{r^2} e^{i \kappa r} y_{kl}(r) y_{k'l'}(r) dr \;+\; c.c. \qquad (20)$$

The ESC-type scheme presented by Sil et al [21] was designed for the numerical evaluation of the integrals of eq. (19). Nevertheless, it is straightforward to extend it to the more complex integrals of the type (20), which is what we did. We have found that the results thus obtained have high numerical accuracy. This has been tested by first evaluating the above integral for two radii, say $r_1$ and $r_2$, subtracting the two values obtained, and then comparing the result with the numerical calculation of the integral evaluated numerically from $r_1$ to $r_2$ using a small step. The results of the two types of calculation agree to up four significant digits.

Again, away from threshold the main contribution comes from integrands containing $e^{\pm i(\phi_{kl}(r)-\phi_{k'l'}(r) \pm \kappa r)}$. Moreover, at values of the photon wavenumber, $\kappa \geq k+k'$, the integrand $e^{\pm i(\phi_{kl}(r)+\phi_{k'l'}(r) - \kappa r)}$ also makes an important contribution and results in a qualitative different integral.

Using the division of the operator of eq. (14), we write the outer-region contribution of the integrals (17) as,

$$\mathcal{E}_{kl;k'l'}^{\kappa\ell}(r_0) = \frac{2\ell+1}{\kappa}\left[ A_\ell S_{kl;k'l'}(r_0) + G_{\kappa\ell;kl;k'l'}^{(-2)}(r_0) \right] \tag{21}$$

where $S_{kl;k'l'}$ is the overlap between the scattering functions,

$$S_{kl;k'l'}(r_0) \equiv \int_{r_0}^{\infty} y_{kl}(r)y_{k'l'}(r)dr \tag{22}$$

As shown in Appendix A of [19], the overlap (22) is related to the *acceleration* integral (19) by the formula,

$$S_{kl;k'l'}(r_o) = P\frac{1}{(\varepsilon-\varepsilon')}[C_{ll'}S_{kl;k'l'}^{(-2)}(r_o) + W_{kl;k'l'}(r_o)]$$
$$+ \cos(\delta_{kl}-\delta_{kl'})\delta(\varepsilon-\varepsilon') \tag{23}$$

$\delta(\varepsilon-\varepsilon')$ is the delta function added to a principal value ($P$) integrand. $W_{kl;k'l'}(r_0)$ is a surface term defined as,

$$W_{kl;k'l'}(r) = \frac{1}{2}\left[ y_{k'l'}(r)\frac{d}{dr}y_{kl}(r) - y_{kl}(r)\frac{d}{dr}y_{k'l'}(r) \right] \tag{24}$$

Moreover,

$$C_{ll'} = \frac{1}{2}l(l+1) - \frac{1}{2}l'(l'+1) \tag{25}$$

The matrix element (17) is the sum of the two contributions, *inner* and *outer*. Away from the point $\varepsilon = \varepsilon'$, we can neglect the principal value and the delta function of eq. (23) and write,

$$\mathcal{E}_{kl;k'l'}^{\kappa\ell} = \frac{2\ell+1}{\kappa}\left\{\int\limits_0^{r_0} I_\ell^{-1}(\kappa r)y_{kl}(r)y_{k'l'}(r)dr + G_{\kappa\ell;kl;k'l'}^{(-2)}(r_0)\right.$$
$$\left. + A_\ell\frac{1}{(\varepsilon-\varepsilon')}[C_{ll'}S_{kl;k'l'}^{(-2)}(r_0) + W_{kl;k'l'}(r_0)]\right\} \tag{26}$$

The expression in brackets is independent of $\kappa$.

In eq. (26), the non-singular part, which is the sum of the first two terms, is separated from the singular one. This separation, although necessary for the calculation of the principal value integration in the region about the singularity, has a serious drawback, as it leads to wildly oscillating functions. Evidently, this feature renders them numerically inconvenient and troublesome as integrands. This is due to the fact that integrals of different functions have been cut-off at $r_0$.

Using eq. (14), we can rewrite eq. (26) as,

$$\mathcal{E}_{kl;k'l'}^{\kappa\ell} = \frac{2\ell+1}{\kappa}\left\{\int\limits_0^{r_0} J_\ell^{-1}(\kappa r)y_{kl}(r)y_{k'l'}(r)dr + G_{\kappa\ell;kl;k'l'}^{(-2)}(r_0)\right.$$
$$\left. + A_\ell\left[\int\limits_0^{r_0} y_{kl}(r)y_{k'l'}(r)dr + \frac{1}{(\varepsilon-\varepsilon')}[C_{ll'}S_{kl;k'l'}^{(-2)}(r_0) + W_{kl;k'l'}(r_0)]\right]\right\} \tag{27}$$

As stated above, $G^{(-2)}(r_0)$ is the outer-part contribution of the operator $J_\ell^{-1}(z)$. Therefore, the sum of the first two terms computes the matrix element of this operator. As such, it is independent of $r_0$ and exhibits no oscillations. The same is true for the sum of the last two terms. In order to make these features apparent, we define the function,

$$C_{ll'}T_{kl;k'l'}(r_0) = (\varepsilon-\varepsilon')\int\limits_0^{r_0} y_{kl}(r)y_{k'l'}(r)dr + W_{kl;k'l'}(r_0) \tag{28}$$

Then, we may write

$$\mathcal{E}_{kl;k'l'}^{\kappa\ell} = \frac{2\ell+1}{\kappa}\left\{J_{\kappa\ell;klk'l'}^{(-1)}(0) + A_\ell\frac{1}{(\varepsilon-\varepsilon')}C_{ll'}(T_{kl;k'l'}(r_0) + S_{kl;k'l'}^{(-2)}(r_0))\right\} \tag{29}$$

This form of the matrix element is convenient when integrating over the energy singularity, since one requires a well-behaved function in the principal-value integral. (The $\delta$-function must also be included in the integration).

For hydrogenic functions, where no core orbitals are present, we obtain the simple result

$$T_{kl;k'l'}(r_0) + S_{kl;k'l'}^{(-2)}(r_0) = S_{kl;k'l'}^{(-2)}(0) \qquad (30)$$

The zero in the function of the right-hand side indicates that the *acceleration* integral, eq. (19), now extends from zero to infinity. (See Appendix A of [19]). Excluding the overall factor $\kappa^{-1}$, the singular term has no dependence on $\kappa$ and has a rather simple functional form. On the contrary, the first term, which is the non-singular part of the operator, contains all the numerical complexity of $\boldsymbol{O}_E(\kappa, \vec{r})$.

At this point, it is appropriate to insert a comment regarding a previously proposed by us simplified model of the electric operator [19]. This model consists of replacing the quantity $I_\ell^{-1}$ by its small-$r$ limit, up to a finite distance $R$, while neglecting the quantity $G^{(-2)}(R)$ altogether. Within this model, the value of $R$ is determined to be $\frac{3\pi}{4\kappa}$, a result that was derived for the $s \to p$ transition. However, integrals which extend up to a finite distance and where at least one scattering orbital is involved, become highly oscillatory as a function of energy, and hence are not suitable as kernels, especially over singularities. The method of the accurate calculation of $G^{(-2)}(r_0)$ described above overcomes this problem completely.

Returning to the discussion of the operator $\boldsymbol{O}_E(\kappa, \vec{r})$, we note that the separation into the two terms of eq. (29) is desirable mainly in an energy region about the singularity $k = k'$. As it is shown by the numerical calculations below, this region is uniquely defined by the photon wavenumber, $\kappa$, as $k - \kappa \le k' \le k + \kappa$. In this region the asymptotic momentum is conserved. As we shall see, the behavior of the matrix elements inside this region is very different from that in the outside.

## VI.  THE MATRIX ELEMENTS OF THE FULL ELECTRIC FIELD OPERATOR

In the case of dipole selection rule (E1), $l' \to l \pm 1$, $m' = m \pm 1$, or quadrupole selection rule (E2), $l' \to l \pm 2$, $m' = m \pm 1$, the matrix element of $\boldsymbol{O}_E(\kappa, \vec{r})$, $\mathcal{E}_{klmk'l'm'}^{\kappa}$, is formed by the sum of the partial radial matrix elements, eq. (27). The summation runs from $\ell = |l - l'|$ to $l + l'$ in steps of 2, each radial integral multiplied by an angular factor $< Y_l^m \mid \Theta_\ell(\theta, \phi) \mid Y_{l'}^{m'} >$, which is computable in terms of 3-$j$ symbols [16,17].

For the dipole selection rules, in order to compare with the corresponding radial matrix element of the *length* EDA operator, we divide the sum by the angular factor $< Y_l^m \mid \Theta_1(\theta, \phi) \mid Y_{l'}^{m'} >$ of the first term, obtaining the *radial* matrix element $\bar{\mathcal{E}}_{klmk'l'm'}^\kappa$:

$$\bar{\mathcal{E}}_{klm;k'l'm'}^\kappa = \frac{1}{\kappa}\left\{ J_{\kappa;klm;k'l'm'}^{-1}(0) + A_{lml'm'}\frac{1}{(\varepsilon - \varepsilon')}C_{ll'}(T_{kl;k'l'}(r_0) + S_{kl;k'l'}^{(-2)}(r_0)) \right\} \quad (31)$$

This is the quantity to be compared with the radial dipole $D_{kl;k'l'}$ matrix element, whose form we shall discuss in the next section.

The non-singular term $J_{\kappa 1;k0k'1}^{-1}$ of eq. (31) is shown for $\kappa = 0.02$ a.u., in **Figures 1-4**, for the transitions $kp \rightarrow k'd$ and $kp \rightarrow k's$ obeying the E1 selection rule, and in **Figures 5, 6** for the transition $kp \rightarrow k'f$ obeying the E2 selection rule. These are shown for the Neon as well as for the Hydrogen atoms. We stress that, although the hydrogenic eigenfunctions are known analytically, the calculations here are numerical. As expected, the differences between the Ne and the H *f-f* matrix elements are due to the presence of the core orbitals of the Ne$^+$ 2$s$-hole state.

We note that the operator resulting from the EQA ($\sim r^2$) has not been employed for the calculation of *f-f* matrix elements. Free-free matrix elements of an unbounded operator such as ($\sim r^2$), present extreme mathematical difficulties and have not been calculated yet using energy-normalized scattering functions.

The derivative of the $J_{\kappa 1;k0k'1}^{-1}$ function, shown in the inset figures, reveals sharp inflection points at the wavenumbers $k' = k \pm \kappa$. *The existence of such a region is a central result of the mathematical analysis of the f-f matrix element of the full electric operator.* Inside this region, where the asymptotic momentum is conserved, i.e., $\vec{k}' = \vec{k} + \vec{\kappa}$, the behavior of the function is markedly different from that on the outside.

The function $T_{kl;k'l'}(r_0) + S_{kl;k'l'}^{(-2)}(r_0)$ appearing at the numerator of the singular part in eqs. (29) and (31) is depicted in **Figures 7, 8** for the transitions $kp \rightarrow k'd$ and $kp \rightarrow k's$ respectively for Neon, together with the quantity $S^{(-2)}_{kl;k'l'}(0)$ for Hydrogen. In the case of Hydrogen, these two functions coincide and are equal to $S^{(-2)}_{kl;k'l'}(0)$, whereas in the case of Neon they differ quantitatively. In addition, also shown is the corresponding $F_{kl;k'l'}(r_0) + S^{(-2)}_{kl;k'l'}(r_0)$ function for the *length* EDA for Ne, a case which is discussed in the next section. In **Figure 9**, the same quantities are depicted for the E2 transition, $kp \rightarrow k'f$. Here, the quantities $T_{kl;k'l'}(r_0) + S^{(-2)}_{kl;k'l'}(r_0)$ for Neon and $S^{(-2)}_{kl;k'l'}(0)$ for Hydrogen, differ qualitatively as well as quantitatively. In these graphs, the discontinuity of the derivative of the numerator of the singular part at $k = k'$ is clearly visible. It is known from the hydrogenic case that it is due to a logarithmic singularity [7, 8]. Thus, integration about this point must be done with extreme care.

## VII. THE ELECTRIC DIPOLE OPERATOR *f-f* MATRIX ELEMENTS

In comparison to eq. (26), in the outer region the matrix element for the dipole *length* operator involves a more complicated surface term, $X$, in addition to $W$. This is a derivative of the delta function and a second-order principal value integral [19,20]. Away from the point $\varepsilon = \varepsilon'$, it is given by the expression,

$$
D_{kl;k'l'} = \int_0^{r_o} r\, y_{kl}(r) y_{k'l'}(r) dr \\
+ \frac{1}{(\varepsilon - \varepsilon')^2}[Z S^{(-2)}_{kl;k'l'}(r_o) + X_{kl;k'l'}(r_o) + (\varepsilon - \varepsilon') r_o W_{kl;k'l'}(r_o)]
$$
(32)

where Z is the nuclear charge.

The inner and outer parts of the dipole matrix elements as functions of $\varepsilon'$ at fixed energy $\varepsilon$, are wildly oscillating functions. Again, we define a function

$$
Z F_{kl;k'l'}(r_o) = (\varepsilon - \varepsilon')^2 \int_0^{r_o} r\, y_{kl}(r) y_{k'l'}(r) dr + [X_{kl;k'l'}(r_o) + (\varepsilon - \varepsilon') r_o W_{kl;k'l'}(r_o)] \quad (33)
$$

Then, we rewrite eq. (32) as

$$D_{kl;k'l'} = \frac{Z}{(\varepsilon - \varepsilon')^2} [F_{kl;k'l'}(r_o) + S^{(-2)}_{kl;k'l'}(r_o)] \qquad (34)$$

For hydrogenic functions, where no core orbitals are present, the following relation holds [19],

$$F_{kl;k'l'}(r_0) + S^{(-2)}_{kl;k'l'}(r_0) = S^{(-2)}_{kl;k'l'}(0) \qquad (35)$$

Thus, upon substitution into eq. (34), we obtain, in a totally different way, the well-known result for hydrogen, originally computed analytically by Gordon [25].

## VIII.  DETAILS OF THE CALCULATION

The radius $r_0$ defining the inner region depends on the value of $\kappa$. It is chosen so that the continued fraction which is employed for the computation of the function $si(\kappa r)$ of eq. (13), (also Appendices A and B), converges in the complex plane.

The spatial calculation of the matrix elements is done in terms of wavenumbers. Thus, the first scattering orbital, $kp$, is held fixed at the wavenumber $k = 1.0$ a.u., while the second one takes values in the interval $k' = [0.8, 1.2]$ a.u., in steps of $10^{-3}$ a.u.. The calculation of the matrix element of $\boldsymbol{O}_E(\kappa, \vec{r})$ is performed for $\kappa = 0.02$ a.u.. This is in the extreme ultraviolet region of the spectrum. The EDA case, which corresponds to $\kappa = 0$, is also computed.

We examine the transitions from $kp$ to $kd$ and to $ks$ scattering orbitals, which obey E1 selection rules, and the transition to the $kf$ orbital, which obeys the E2 rule, for Hydrogen and for Neon. In both cases, the scattering orbitals at relatively small-$r$ values are numerical. For Ne, they are calculated in the fixed-core HF approximation. They extend up to the radius $r_c = 20$ a.u., where the core orbitals are practically zero. Beyond this radius and up to the radius $r_0$ defining the inner region, their numerical values can be calculated either by a numerical solution of the resulting hydrogenic equation, or by a phase-shifted WKB form [19]. For the Hydrogen atom where no core orbitals are present, the intermediate radius $r_c$ is not used since it has no meaning.

## IX. COMPARISON OF THE MATRIX ELEMENTS OF THE TWO OPERATORS

The expressions of the matrix elements $\bar{\mathcal{E}}^{\kappa}_{klmk'l'm'}$, and those of the corresponding EDA, $D_{kl;k'l'}$, look quite different. Yet, they give results that are very close in magnitude, except when obtained inside the critical region of extension $\kappa$ about the singular point $k = k'$. (See **Figures 10, 11**). It appears that, away from this region, the non-singular term of eq. (31) nearly cancels the singular one, leaving the dipole matrix element of eq. (32) as a residue.

For example, for the transition $kp \rightarrow k's$ in Hydrogen, the non-singular part of the matrix element outside the critical region, is well approximated by a function of the form

$$f(k,k') = \frac{Z - a(\varepsilon - \varepsilon')}{(\varepsilon - \varepsilon')^2} S^{(-2)}_{kl;k'l'}(0) \quad , \qquad a = \frac{3\pi}{4\kappa} C_{01} \qquad (36)$$

In this simple case, the term $a S^{(-2)}_{kl;k'l'}(0)/(\varepsilon - \varepsilon')$ cancels the second term of the right-hand of eq. (31), leaving the EDA result of Gordon [25], eqs. (34,35).

The implication of the previous analysis is the following:

*The exceptional numerical features of the matrix elements of $\boldsymbol{O}_E(\kappa, \vec{r})$ result principally from the non-singular term, and are confined in the region $k - \kappa \le k' \le k + \kappa$.*

We elaborate: In actual calculations, the transition matrix elements containing pairs of scattering orbitals, enter as integrands in integrals over wavenumber (or energy). These integrals are the sum of contributions from three regions:

$$\int_{0}^{k-\kappa} dk', \quad \int_{k-\kappa}^{k+\kappa} dk', \quad \int_{k+\kappa}^{\infty} dk'$$

As we have shown, only in the middle integral does the result from the EDA become very different from that of the full electric operator in the calculation of the *f-f* matrix elements. Its range of integration depends on the photon wavenumber, $\kappa = \omega / c$. Hence, in the low-frequency region of the spectrum, the range where the *f-f* matrix elements of $\boldsymbol{O}_E(\kappa, \vec{r})$ differ significantly from those obtained from the EDA, is very narrow.

Given the above results and analysis, future attempts at the non-perturbative solution of the METDSE with energy-normalized scattering functions ought to take into account the fact that the singular behavior of the matrix elements of $\boldsymbol{O}_E(\kappa, \vec{r})$ is $(\varepsilon - \varepsilon')^{-1}$, which is simpler than that of the EDA, whose singularity goes like $(\varepsilon - \varepsilon')^{-2}$. At this point we mention that the constant $A_{lml'm'}$ in eq. (31) becomes zero for $l + m = odd$ values, thereby eliminating the singularity. This case is not examined in the present work, where the initial excitation from the 2s-orbital leads to $kp_1$ by the use of the dipole selection rule, i.e. to a level for which $l + m = 2$.

## X. SYNOPSIS

Given the continuing experimental advances in the generation and spectroscopic use of strong and/or ultra-short electromagnetic pulses and of related pump-probe techniques, e.g., [26,27], the scope of theoretical N-electron atomic and molecular physics has been broadened significantly, now incorporating the possibility of many challenging problems of *time-resolved many-electron physics*. In this realm, the fundamental requirement is the possibility of solving the *time-dependent Schrödinger equation* nonperturbatively, for initial states with arbitrary electronic structures. To this purpose, we have proposed and applied the *state-specific expansion approach* [1,2,5]. Its implementation requires the computation of matrix elements of the operator that describes the coupling between energy-normalized scattering states to the external electromagnetic field.

For free-free transition matrix elements, which are the object of the present study, the standard notion of 'atomic dimensions' cannot be invoked for the a priori justification of the textbook *multipole* expansion of the *minimal- coupling* interaction derived, e.g., from the Taylor expansion, $e^{i\vec{k}\vec{r}} = 1 + i\vec{k}\vec{r} + \ldots$ . Instead, the present theory has focused on choices of selection rules, (according to experimental conditions and to the choice of initial and final scattering states), of parity, spin and angular momentum change, e.g., $\Delta j = 0, \pm 1, \pm 2, \ldots$ . The examples used were the electric dipole and the electric quadrupole selection rules. Depending on this choice, although the principal contribution comes, of course, from the electric dipole operator (e.g., parity changes and $\Delta j = \pm 1$ ), and, to a much smaller degree, from the electric quadrupole operator, (parity does not change and $\Delta j = \pm 2$ ), etc., the total interaction also causes contributions from higher order terms. (For example, for $\Delta j = \pm 1$ , the first correction to the free-free electric dipole matrix element comes from the electric octopole).

We presented the results of mathematical analysis and of computations of matrix elements of $\boldsymbol{O}_E(\kappa, \vec{r})$ between state-specific, energy-normalized, numerical scattering wavefunctions for Hydrogen (whose matrix elements can be tested using analytic eigenfunctions) and for Neon, *on*- and, especially, *off-resonance,* $(\varepsilon - \varepsilon' \neq \hbar\omega)$ . Emphasis was given on the identification of singularities in these *f-f* matrix elements as $\varepsilon' \to \varepsilon$ , and on understanding their mathematical nature, so as to devise practical computational schemes involving energy-normalized, numerical scattering orbitals.

As regards numerical results, the quantities in eq. (31), (full electric operator), and in eq. (34), (EDA), are plotted in **Figs. 1-9**. The non-singular term $J_{\kappa 1; k0k'1}^{-1}$ of eq.(31) is shown for $\kappa = 0.02$ a.u., in **Figs. 1-4** for the transitions $kp \to k'd$ and $kp \to k's$ obeying the E1 selection rule, and in **Figs. 5, 6** for the transition $kp \to k'f$ obeying the E2 selection rule. These are shown for Neon as well as for Hydrogen.

The matrix elements of the full electric operator and of the *length* EDA differ qualitatively in the region $k - \kappa \leq k' \leq k + \kappa$ , where $\kappa$ is the photon wavenumber. (**Figures 10, 11**). In particular, the singularity, as $\varepsilon' \to \varepsilon$ , of the matrix elements of $\boldsymbol{O}_E(\kappa, \vec{r})$, eq. (31), is of the type $(\varepsilon - \varepsilon')^{-1}$, whilst that of the EDA, eq. (34), is of the type $(\varepsilon - \varepsilon')^{-2}$.

In conjunction with the herein analysis, these facts indicate that, when *numerical* channel-dependent scattering functions are used, as is necessary when dealing with arbitrary electronic structures of atoms, the matrix elements of the full electric operator are "easier" to handle numerically in the regions of the $\varepsilon = \varepsilon'$ singularities, than those involving the electric dipole or quadrupole (and higher) operators.

Finally, a comment on the overall size of the calculation is relevant: For nonperturbative, strong-field calculations, whether in the dipole or in the full-interaction description, the number of angular momentum states in the continuous spectrum must be large. Many multipoles may become important for the calculation of free-free matrix elements, using the same large basis. The present results and arguments show that the matrix elements using the full operator present fewer difficulties.

## APPENDIX A: ELECTRIC OPERATOR FOR ARGUMENTS $|z| \geq 1$

We shall show below that, for arguments $|z| \geq 1$, the radial integrals of the electric operator given by eq. (11) can be expressed in terms of functions of the general form,

$$\frac{Q(z)}{z^2} e^{iz} , \qquad \lim_{z \to \infty} Q(z) \to 1 \tag{A1}$$

where $z$ is a complex variable.

We first note that the spherical Bessel functions entering eq. (11) can be expressed in terms of the spherical Hankel functions [18] as,

$$j_\ell(z) = \frac{1}{2}[h_\ell^{(1)}(z) + h_\ell^{(2)}(z)] \tag{A2}$$

where

$$h_\ell^{(1)}(z) = (-i)^\ell \frac{e^{iz}}{iz} \sum_{k=0}^{\ell} \frac{(\ell+k)!}{k!(\ell-k)!} \frac{1}{(-2iz)^k} \tag{A3}$$

$$h_\ell^{(2)}(z) = i^\ell \frac{e^{-iz}}{(-iz)} \sum_{k=0}^{\ell} \frac{(\ell+k)!}{k!(\ell-k)!} \frac{1}{(2iz)^k} \qquad (A4)$$

Thus, the functions $j_\ell(z)/z$ in eq.(11), are of the general form (A1).

We then examine the function $I_1^{-1}(z)$ of eq. (10a) which, in this region, is conveniently written as $Si(z) \equiv \frac{\pi}{2} + si(z)$. The later function, $si(z)$, can be expressed in terms of the exponential integral function $E_1(z)$ as

$$si(z) = \frac{1}{2i}[E_1(iz) - E_1(-iz)] \qquad (A5)$$

$E_1$ is the first member of the family of exponential integrals defined in [18], (see also Appendix B). For $|z| \gtrsim 1$, they are conveniently represented in terms of a continued fraction, $F_n(z)$, times an exponential:

$$E_n(z) = F_n(z)e^{-z} \qquad (A6)$$

From the asymptotic expression of $E_n(z)$, for $|z| \to \infty$ we get $F_n \to z^{-1}$ (see Appendix B), so we can write

$$E_n(z) = \overline{F}_n(z)\frac{e^{-z}}{z} \qquad \text{with} \qquad \overline{F}_n(z) \to 1 \qquad (A7)$$

The $z^{-1}e^{-z}$ term, is canceled by the similar term of $j_1(z) = -\frac{\cos z}{z} + \frac{\sin z}{z^2}$ in eq.(10a). Indeed, one can easily show from the definition (B1) that,

$$E_1(z) - \frac{e^{-z}}{z} = -\frac{1}{z}E_2(z) \qquad (A8)$$

and from (A7) that,

$$\overline{F}_1(z) - 1 = -\frac{1}{z}\overline{F}_2(z) \ . \qquad (A9)$$

Applying these formulas we get,

$$\begin{aligned} si(z) - j_1(z) = &-\frac{1}{2z}[\overline{F}_1(-iz)e^{iz} + \overline{F}_1(iz)e^{-iz}] \\ &+ \frac{1}{2z}(e^{iz} + e^{-iz}) - \frac{1}{2iz^2}(e^{iz} - e^{-iz}) \end{aligned} \qquad (A10)$$

and finally,

$$\frac{1}{2}[si(z) - j_1(z)] = -\frac{1}{2iz^2}[\frac{\overline{F}_2(-iz)+1}{2}e^{iz} - \frac{\overline{F}_2(iz)+1}{2}e^{-iz}]$$

$$\tag{A11}$$

$$= -\frac{1}{z^2}\text{Re}[\frac{\overline{F}_2(-iz)+1}{2i}e^{iz}]$$

# APPENDIX B:  THE CONTINUED FRACTION OF THE EXPONENTIAL INTEGRALS

The exponential integral functions

$$E_n(z) = \int_1^\infty \frac{e^{-zt}}{t^n}dt \tag{B1}$$

can be expressed in the form of a continued fraction  multiplied by an exponential

$$E_n(z) = F_n(z)e^{-z} \tag{B2}$$

The continued fraction, $F_n(z)$, is given by [18]

$$F_n(z) = \frac{1}{z+} \frac{n}{1+} \frac{1}{z+} \frac{n+1}{1+} \frac{2}{z+}... \tag{B3}$$

It can be observed that for large values of the argument

$$F_n(z) \to z^{-1} \tag{B4}$$

For $n = 1$, the fraction is programmed in the "Numerical Recipes" of Press et al [28] as part of the code of the sine-integral function. It is expressed in its so-called "even" form, which converges twice as fast.

For the general case, i.e., for any $n$, the function $F_n(z)$ is,

$$F_n(z) = \frac{1}{n+z-} \frac{n}{n+2+z-} \frac{2(n+1)}{n+4+z-}... \frac{(k-1)(n+k-2)}{z+n+2k-2-}... \tag{B5}$$

for $k \geq 2$.

To our knowledge, expression B5 has not been given before. The code of [28] is easily modified accordingly.

**AUTHOR CONTRIBUTION STATEMENT**: All authors contributed equally to the paper.

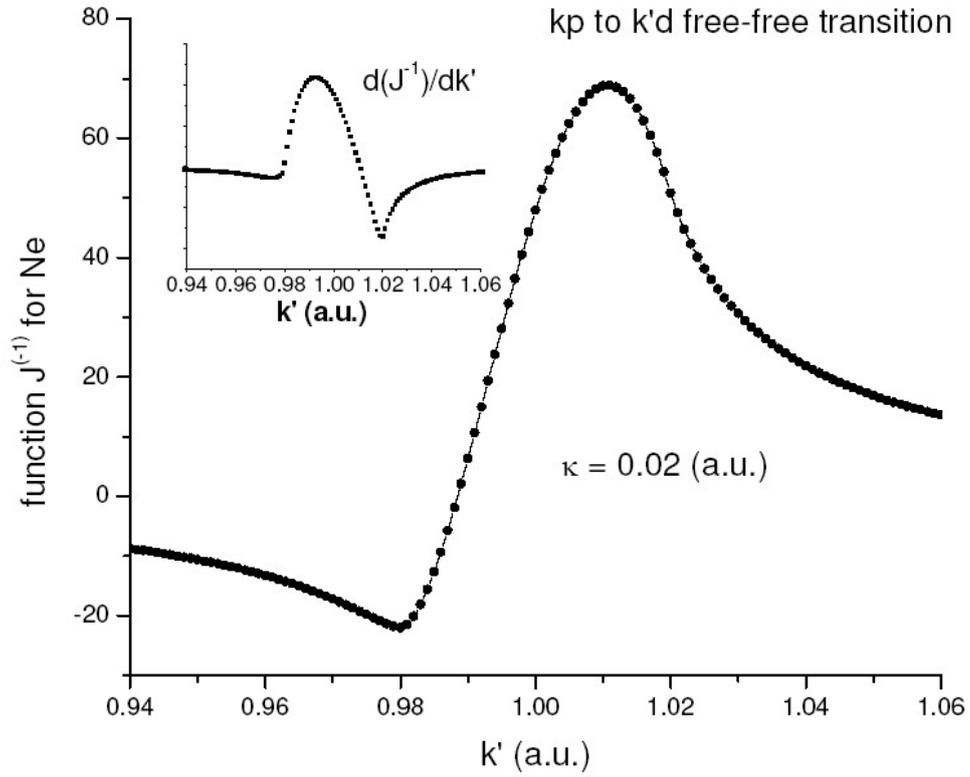

**Figure 1**

The non-singular term, $J^{-1}_{\kappa;klm;k'l'm'}$, of the matrix element of eq. (31) for the transition ($kp \rightarrow k'd$) for Ne. The inset shows the derivative of this function, in arbitrary units. Abrupt changes occur as $k'$ crosses the borders of the region $[k \pm \kappa]$, where $\kappa$ is the photon wavenumber. The figure is drawn for $k = 1.0$ $a.u.$

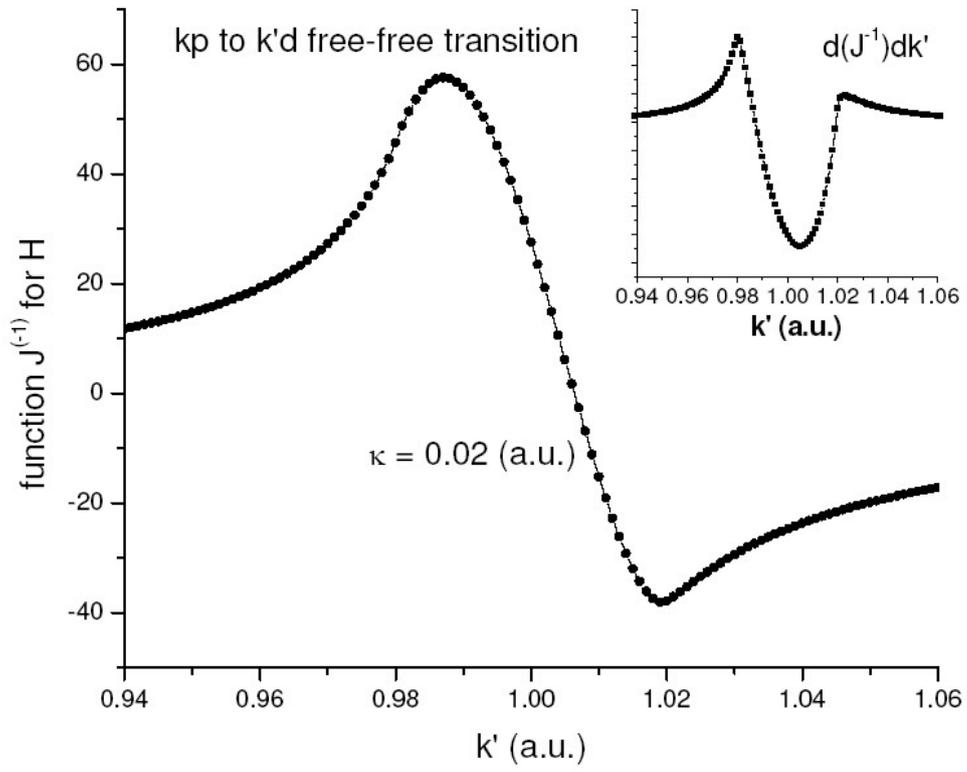

**Figure 2**

As in Fig.1, for Hydrogen.

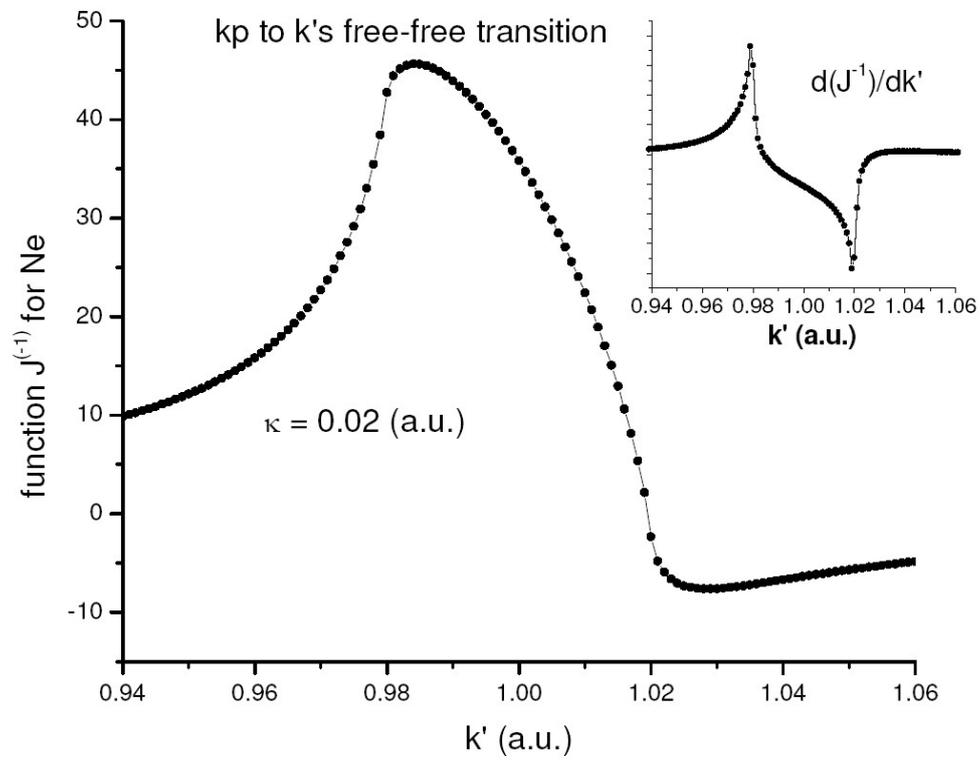

**Figure 3**

As in Fig.1, for the transition $kp \to k's$ .

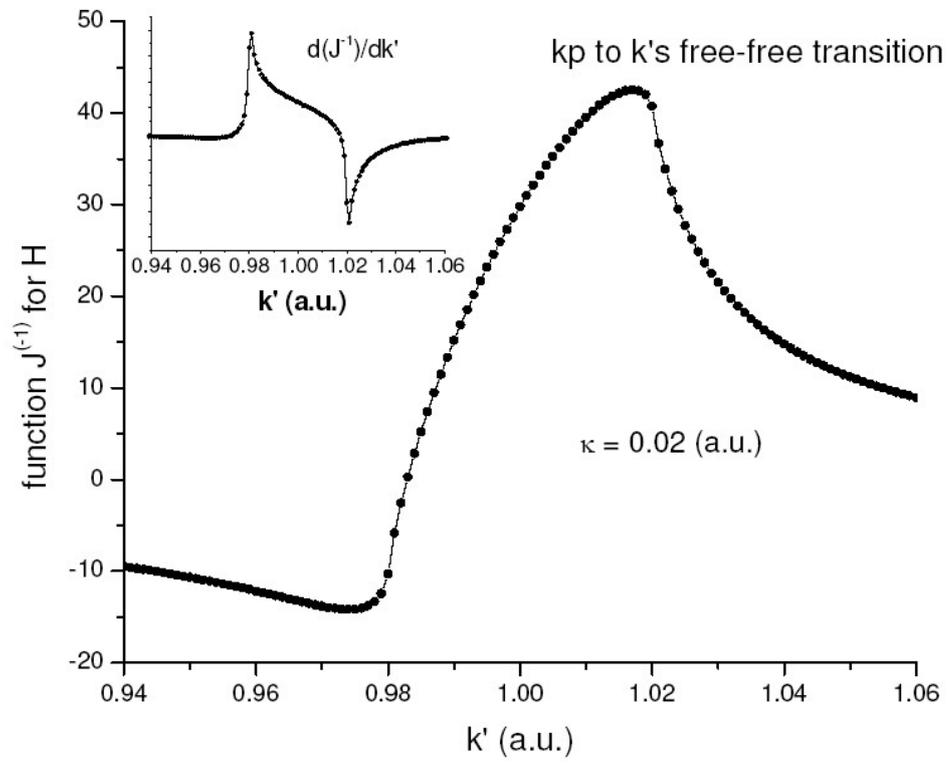

**Figure 4**

As in Fig.3, for Hydrogen.

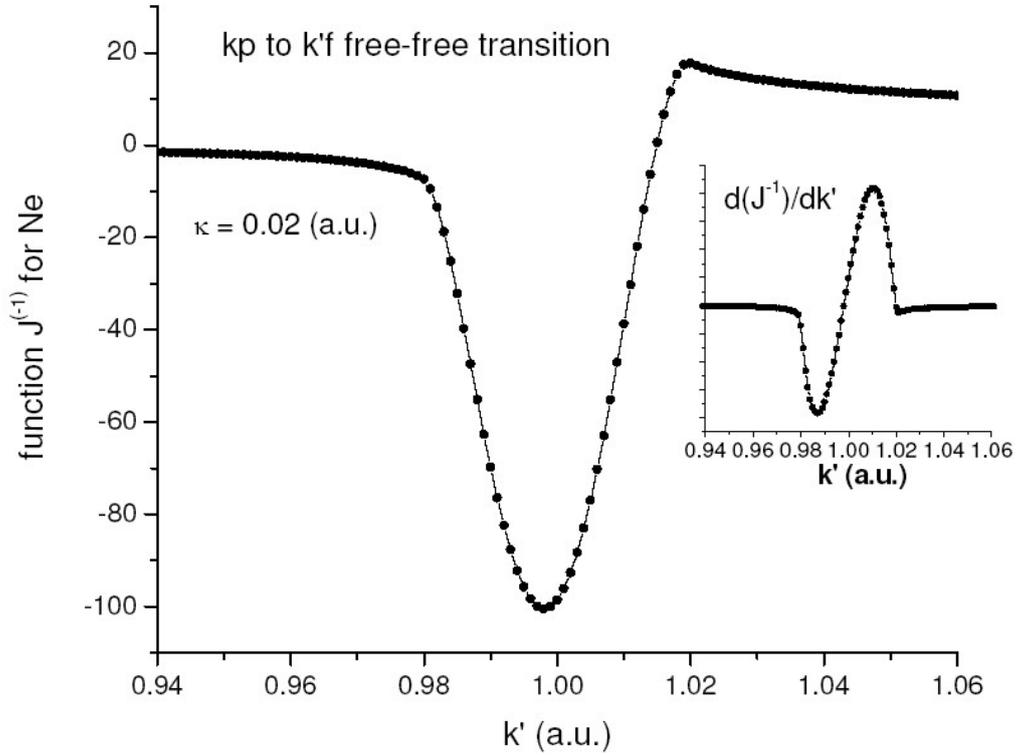

**Figure 5**

The non-singular term, $J^{-1}_{\kappa;klm;k'l'm'}$, of the matrix element of eq. (31) for the transition ($kp \rightarrow k'f$) for Ne. Abrupt changes occur as $k'$ crosses the borders of the region $[k \pm \kappa]$. Note that outside of this region, the values of this term are significantly smaller than those of inside. The figure is drawn for $k = 1.0 \ a.u.$

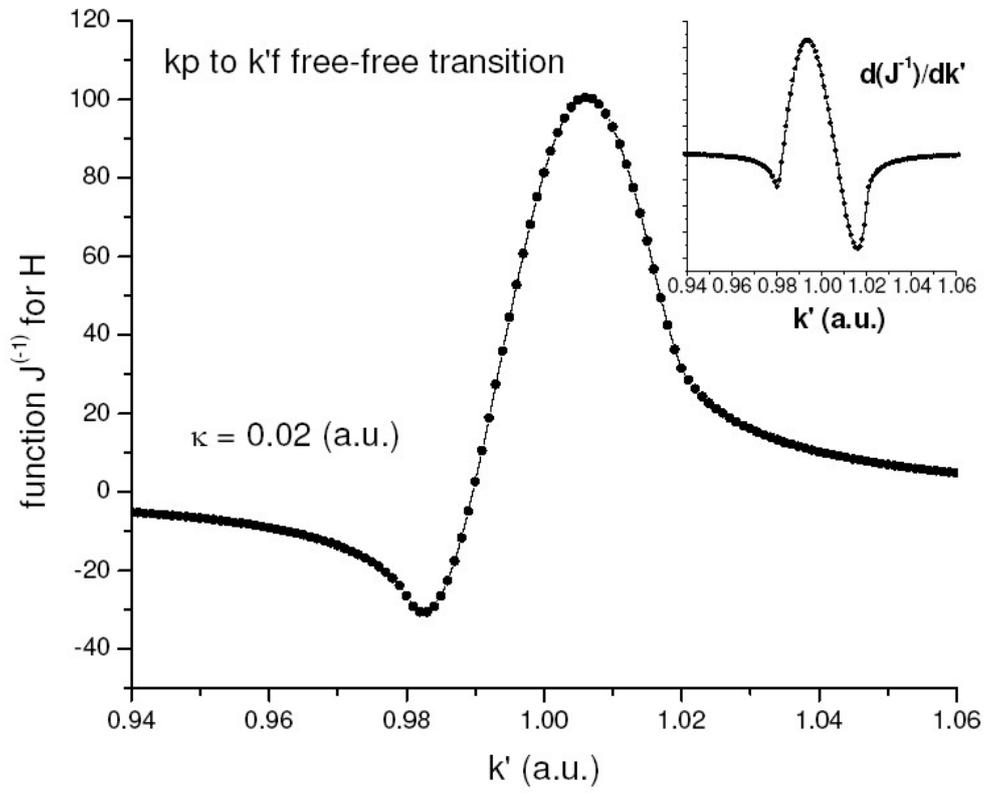

**Figure 6**

As in Fig.5, for Hydrogen.

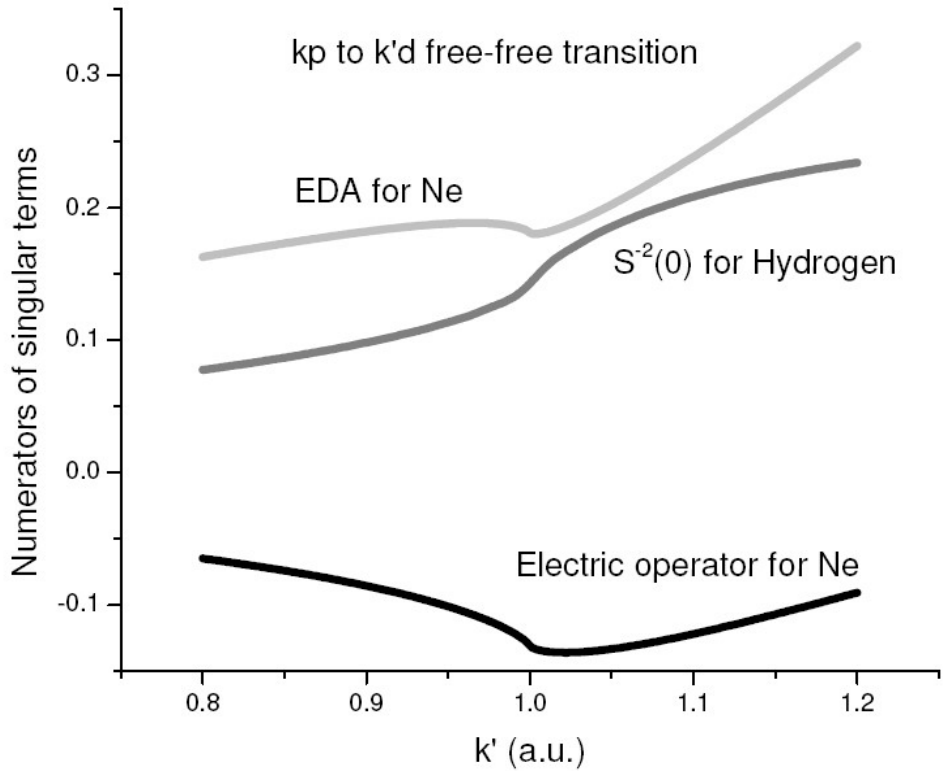

**Figure 7**

The functions of the numerators of the singular parts of the *free-free* transition

($kp \rightarrow k'd$) matrix elements for Neon and Hydrogen (E1 selection rule).

Black line: [T+S] function of eq. (29) for the full electric operator for Ne.

Light gray: [F+S] function of eq. (34) for the EDA for Ne.

Gray line: Special case for Hydrogen, calculated in terms of the *acceleration* integral

(eqs. 30 and 35). In this case, the functions for the full operator and for the EDA are the same.

The difference between the *Ne* results and those of Hydrogen is due to the presence of the core orbitals in $Ne^+$ with a $2s$ hole. Note that the abrupt changes in the three curves signal the logarithmic singularity of the derivative at equal energies, $k' = k$,

see [19]. The figure is drawn for $k = 1.0$ *a.u.*

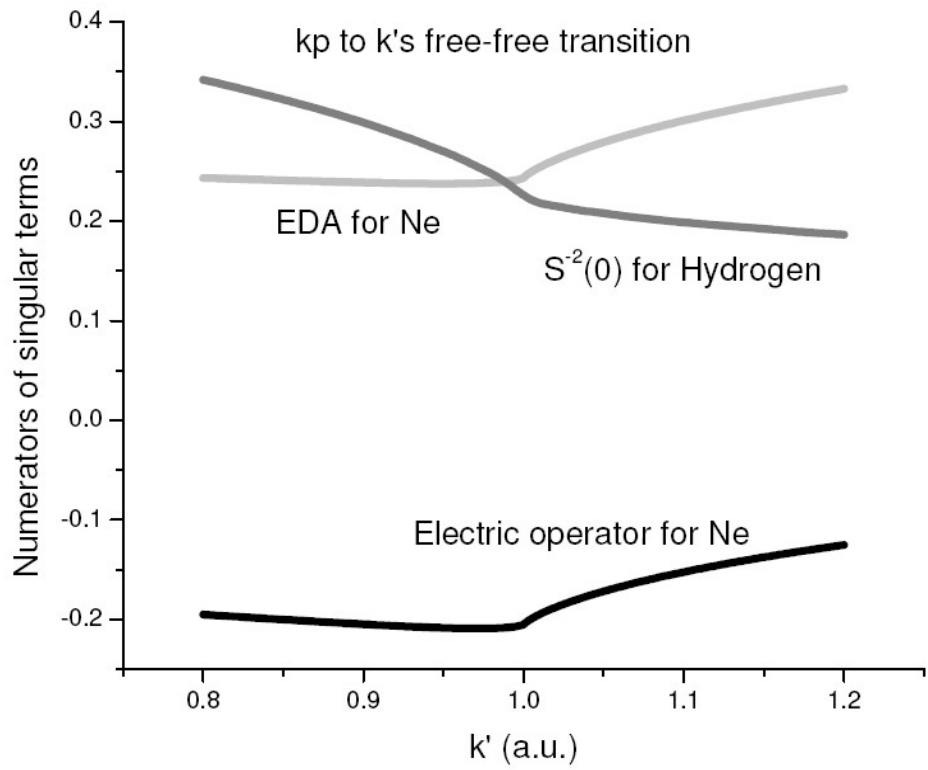

**Figure 8**

As in Fig.7, for the transition $kp \rightarrow k's$.

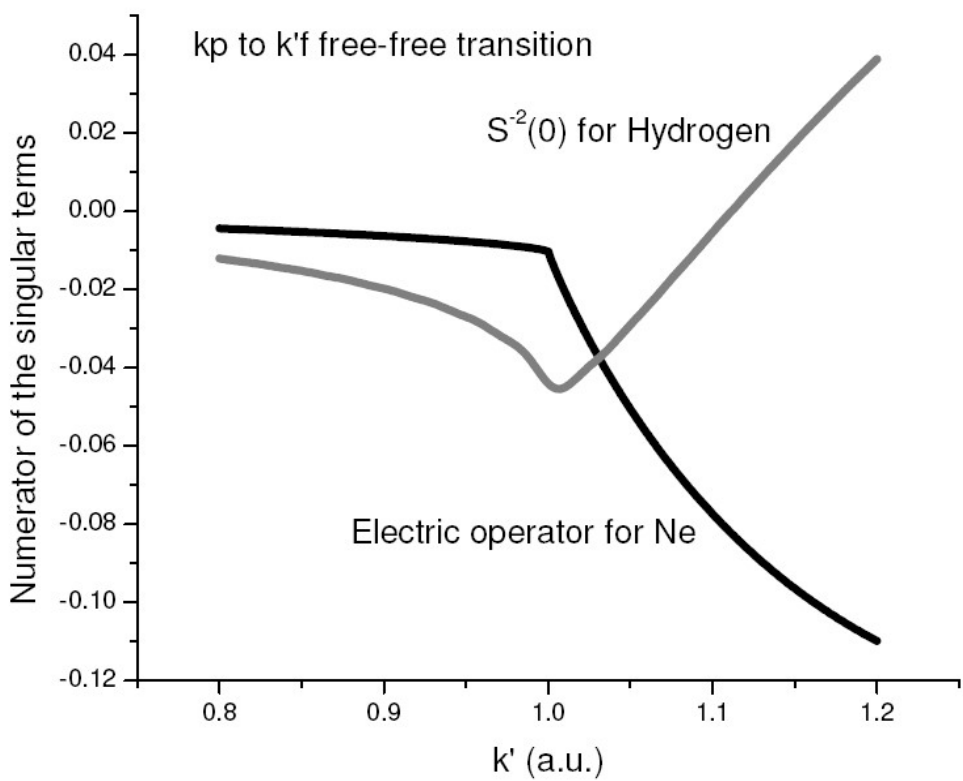

**Figure 9**

The functions of the numerators of the singular parts of the *free-free* matrix elements for the E2 transition $kp \rightarrow k'f$ , for Neon and Hydrogen.

Black line: [T+S] function of eq. (29) for the full electric operator for Ne.

Gray line: Special case for Hydrogen, calculated in terms of the *acceleration* integral (eq. 30). The difference between the Ne results and those of Hydrogen is due to the presence of the core orbitals in $Ne^+$ with a $2s$ hole. The figure is drawn for $k = 1.0$ *a.u.*

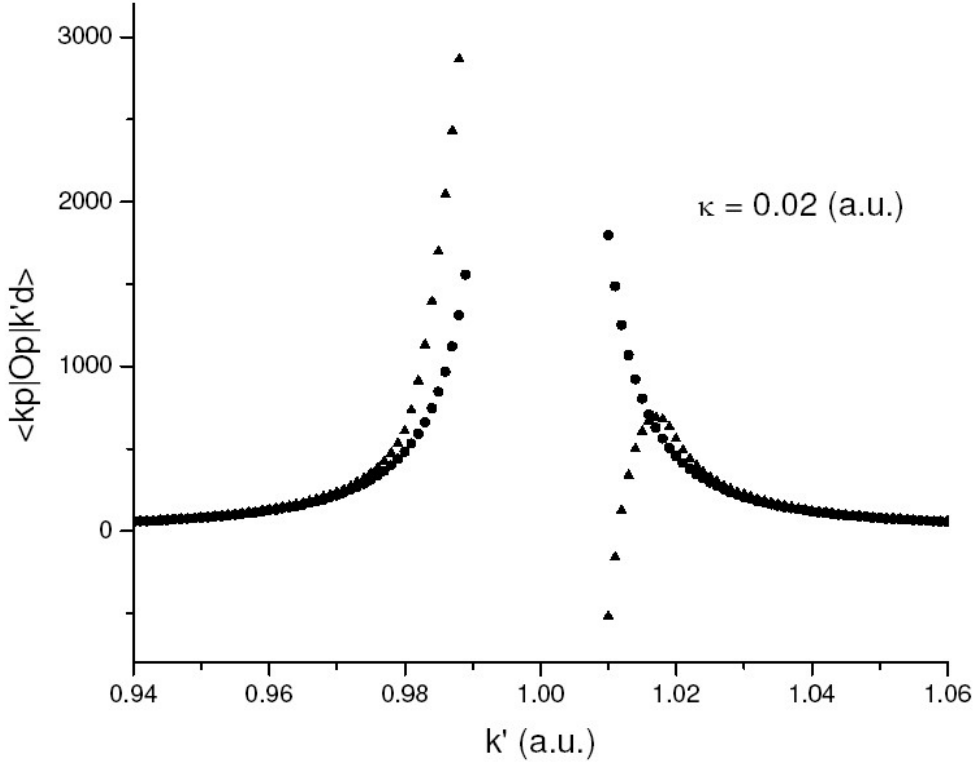

**Figure 10**

The matrix elements $\bar{\mathcal{E}}^{\kappa}_{klm;k'l'm'}$, eq. (31), and $D_{kl;k'l'}$, eq. (34), for the transition

$kp \rightarrow k'd$ for Ne in the region of the singularity ( $k' = k$ ).

Triangles: Full electric operator matrix elements ( $\bar{\mathcal{E}}^{\kappa}_{klm;k'l'm'}$ ).

Circles: EDA matrix elements ( $D_{kl;k'l'}$ ).

The photon energy is 74.6 eV. In the case of the full operator the singularity goes like

$(k' - k)^{-1}$, whilst in the case of the EDA it goes like $(k' - k)^{-2}$.

As discussed in the text, outside the critical region $[k \pm \kappa]$, the matrix elements of the

full electric operator and of its EDA have similar magnitudes. The figure is drawn for

$k = 1.0$ *a.u.*

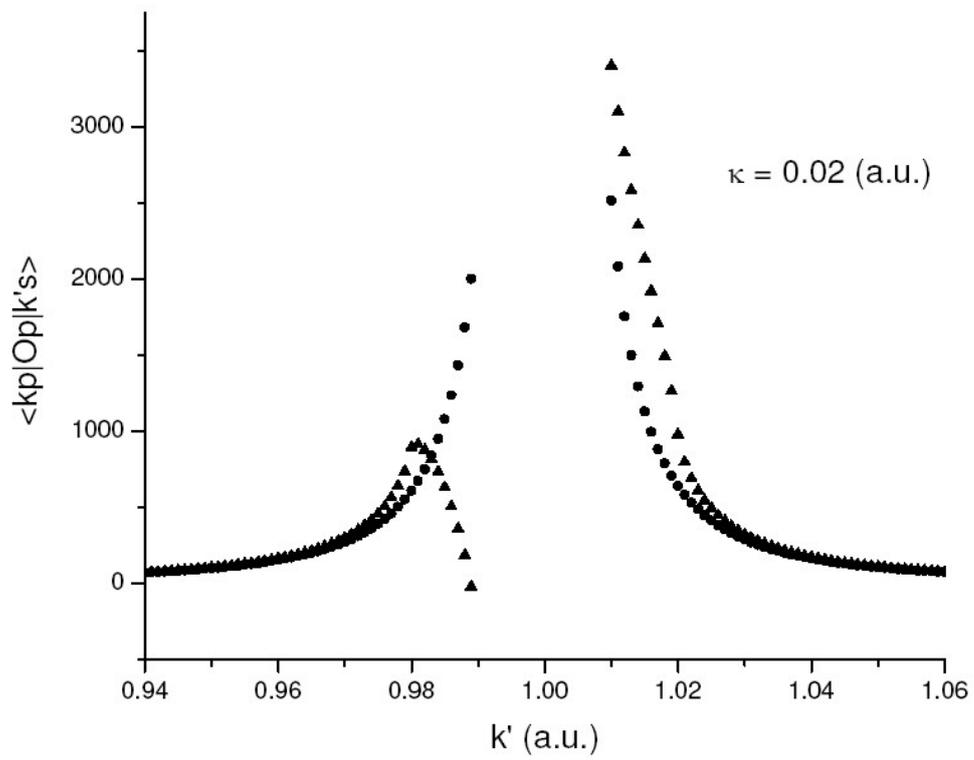

**Figure 11**

As in Fig.10, for the transition $kp \rightarrow k's$ .